%% file: chandler.tex
\documentclass[graybox]{svmult}

\usepackage{mathptmx}       % selects Times Roman as basic font
\usepackage{helvet}         % selects Helvetica as sans-serif font
\usepackage{courier}        % selects Courier as typewriter font
\usepackage{type1cm}        % activate if the above 3 fonts are
\usepackage{makeidx}         % allows index generation
\usepackage{graphicx}        % standard LaTeX graphics tool
\usepackage{multicol}        % used for the two-column index
\usepackage[bottom]{footmisc}% places footnotes at page bottom

\usepackage{bm}
\usepackage{amssymb}

\makeindex     % used for the subject index
\newcommand{\be}{\begin{equation}}
\newcommand{\ee}{\end{equation}}
\newcommand{\bea}{\begin{eqnarray}}
\newcommand{\eea}{\end{eqnarray}}
\newcommand{\ben}{\begin{enumerate}}
\newcommand{\een}{\end{enumerate}}
\newcommand{\bit}{\begin{itemize}}
\newcommand{\eit}{\end{itemize}}

\newcommand{\la}[1]{\label{#1}}

\newcommand{\Eq}[1]{Eq.~(\ref{#1})}

\newcommand{\Fig}[1]{Fig.~\ref{#1}}

\def\vh#1{\hat{\bm{#1}}}							
\newcommand{\vv}[1]{\bm #1}						% 3-vector
\newcommand{\bert}{\raise-0.45mm\hbox{\Large$\Box$}}		%D'Alembertian

\begin{document}

\title*{Chandler wobble: Stochastic and deterministic dynamics}
\author{Alejandro Jenkins}
\institute{Alejandro Jenkins \at Escuela de F\'isica, Universidad de Costa Rica, 11501-2060, San Jos\'e, Costa Rica, and Academia Nacional de Ciencias, 1367-2050, San Jos\'e, Costa Rica, \email{alejandro.jenkins@ucr.ac.cr}}
\maketitle

\abstract*{We propose a model of the Earth's torqueless precession, the ``Chandler wobble'', as a self-oscillation driven by positive feedback between the wobble and the centrifugal deformation of the portion of the Earth's mass contained in circulating fluids.  The wobble may thus run like a heat engine, extracting energy from heat-powered geophysical circulations whose natural periods would otherwise be unrelated to the wobble's observed period of about fourteen months.  This can explain, more plausibly than previous models based on stochastic perturbations or forced resonance, how the wobble is maintained against viscous dissipation.  The self-oscillation is a deterministic process, but stochastic variations in the magnitude and distribution of the circulations may turn off the positive feedback (a Hopf bifurcation), accounting for the occasional extinctions, followed by random phase jumps, seen in the data.  This model may have implications for broader questions about the relation between stochastic and deterministic dynamics in complex systems, and the statistical analysis thereof.}

\abstract{We propose a model of the Earth's torqueless precession, the ``Chandler wobble'', as a self-oscillation driven by positive feedback between the wobble and the centrifugal deformation of the portion of the Earth's mass contained in circulating fluids.  The wobble may thus run like a heat engine, extracting energy from heat-powered geophysical circulations whose natural periods would otherwise be unrelated to the wobble's observed period of about fourteen months.  This can explain, more plausibly than previous models based on stochastic perturbations or forced resonance, how the wobble is maintained against viscous dissipation.  The self-oscillation is a deterministic process, but stochastic variations in the magnitude and distribution of the circulations may turn off the positive feedback (a Hopf bifurcation), accounting for the occasional extinctions, followed by random phase jumps, seen in the data.  This model may have implications for broader questions about the relation between stochastic and deterministic dynamics in complex systems, and the statistical analysis thereof.}

%%%%%%%%%%%%%%%%%%%%%%%%%%%%%%%%%%%%%%%%%%%%%%%%%%%%%%%%%%%%%%%
\section{Introduction}
\label{sec:intro}

According to Euler's theory of the free symmetric top, the axis of rotation precesses about the axis of symmetry, unless both happen to align \cite{Euler}.  Applied to the Earth, this implies that the instantaneous North Pole should describe a circle about the axis of symmetry (perpendicular to the Earth's equatorial bulge), causing a periodic change in the latitude of any fixed geographic location (an effect known as ``polar motion'' or ``variation of latitude'').  The angular frequency of this free precession is
\begin{equation}
\omega_{\rm Eu} = \frac{I_3 - I}{I} \Omega_3 ~,
\la{eq:Euler-f}
\ee
where $I \equiv I_1 = I_2$ and $I_3$ are the Earth's moments of inertia, and $\Omega_3$ is the component of its angular velocity along the axis of symmetry, which is also the principal axis with moment of inertia $I_3$.  Many textbook treatments of rigid body dynamics cover this in detail; see, e.g., \cite{Landau}.  (Some authors refer to the corresponding terrestrial motion as the Earth's ``free nutation''; see \cite{Klein}.)

The value of $I / (I_3 - I)$ can be deduced from the periods and amplitudes of the Earth's {\it forced} precession and nutation, driven by tidal torques \cite{Klein}.  The Eulerian period of the Earth's free precession then comes out to $2 \pi / \omega_{\rm Eu} = 306$ days (about 10 months), but Chandler's careful observations, first published in 1891, established that the actual period is about 14 months \cite{Chandler1,Chandler2,Chandler-NAS}.  This motion is therefore known as the ``Chandler wobble''.  A modern estimate \cite{Wilson} of its angular frequency is
\be
\omega_{\rm Ch} = \frac{2 \pi}{433.0 \pm 1.1 ~ (1 \sigma) ~\hbox{days}} ~.
\la{eq:Chandler}
\ee
In 1892, Newcomb explained the discrepancy between $\omega_{\rm Ch}$ and $\omega_{\rm Eu}$ as a consequence of the Earth not being quite rigid. \cite{Newcomb}

Newcomb also opened a debate, which has raged since, on how the Chandler wobble is maintained against the internal viscosities that would damp it away.  Proposed solutions to this puzzle have, for the most part, relied either on stochastic perturbations to the Earth's mass distribution \cite{Jeffreys1,Jeffreys2}, or on marine and atmospheric circulations that might force the wobble near its resonant $\omega_{\rm Ch}$. \cite{Plag,Gross,Aoyama1,Aoyama2}

Here we propose a new model of the Chandler wobble as a weakly non-linear self-oscillation.  Unlike a resonator, a self-oscillator maintains periodic motion at the expense of a power source with no corresponding periodicity \cite{Andronov,Jenkins}.  A self-oscillation modulates the driving force acting on it, establishing {\it positive feedback} between it and the external power source.  On the many names and guises of self-oscillators, see \cite{Jenkins}.  One striking instance of such behavior was provided by the swaying of London's Millennium Bridge when it opened in 2000, driven by the synchronized lateral motion of the pedestrians in response to the bridge's own oscillation. \cite{Strogatz}

In our model, the Chandler wobble is powered by geophysical fluid circulations whose natural periods are unrelated to $\omega_{\rm Ch}$.  Due to the centrifugal force of the Earth's rotation, the wobble itself can modulate those circulations.  That a dynamical delay in the adjustment of a surrounding flow to a solid's displacement can excite and maintain the solid's vibration is an idea dating back to Airy's conceptualization of the action of the vocal cords \cite{Airy}, based on Willis's pioneering research on the mechanism of the larynx \cite{Willis}.  In the case of the Chandler wobble, variations of the magnitude and geographical distribution of the circulations may turn off the positive feedback.  This suggests an explanation of the wobble's extinctions, which occur rarely and are followed by re-excitation with a random phase jump, without obvious connection to major geophysical events. \cite{Malkin} 

%%%%%%%%%%%%%%%%%%%%%%%%%%%%%%%%%%%%%%%%%%%%%%%%%%%%%%%%%%%%%%%
\section{Precession and deformation}
\label{sec:deformation}

Newcomb's simplified treatment in \cite{Newcomb} of the Earth's deformability and its connection to the observed $\omega_{\rm Ch}$ lends itself to a physically intuitive formulation of our self-oscillatory model.  We therefore begin by summarizing Newcomb's argument.  For a full, rigorous treatment of this subject, see \cite{Klein}.

\begin{figure}[b]
\sidecaption
\includegraphics[width=.6\textwidth]{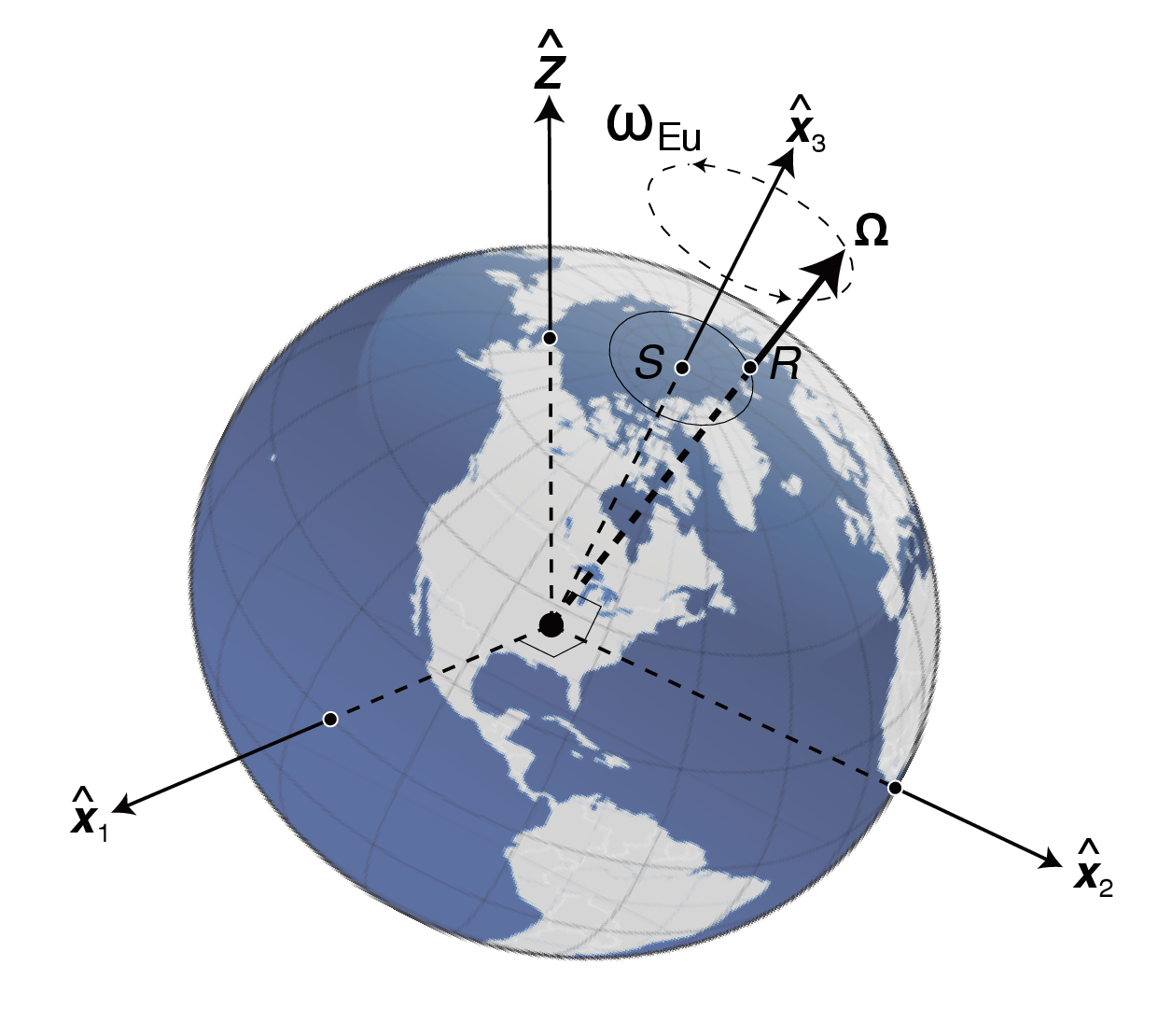}
\caption{Free precession of a rigid Earth.  The principal axes $\vh{x}_{1,2,3}$ define a non-inertial body frame, $\vh{x}_3$ being the Earth's axis of symmetry.  These principal axes move with respect to the inertial space frame (for which only the axis $\vh{Z}$ is shown).  $S$ is the symmetry pole and $R$ the instantaneous North Pole.  In the body frame, $\vv \Omega$ precesses regularly along the circle $SR$, with angular velocity $\omega_{\rm Eu}$.}
\label{fig:Euler} 
\end{figure}

Let $\vv \Omega$ be the Earth's angular velocity and $\vh{x}_3$ be the unit vector along the axis of symmetry, with moment of inertia is $I_3$.  Let $I$ be the value of the moments of inertia along any two axes orthogonal to each other and to $\vh{x}_3$.  Since the Earth is flattened at the poles, $I_3 > I$.  As shown in \Fig{fig:Euler}, for a rigid Earth the instantaneous North Pole $R$ (the intersection of $\vv \Omega$ with the Earth's surface) would describe a circular trajectory (polhode) around the symmetry pole $S$ (the intersection of $\vh{x}_3$ with the Earth's surface), with angular velocity $\omega_{\rm Eu}$ given by \Eq{eq:Euler-f}.

The centrifugal force generated by the planet's rotation implies that a displacement of $R$ tends to deform the planet by shifting its equatorial bulge, causing $S$ to move towards $R$.  If the Earth were simply fluid, this adjustment would be complete and nearly instantaneous, so that $S$ would always coincide with $R$ and there would be no wobble.  For an elastic Earth, the adjustment is partial and the planet may wobble, albeit more slowly than in the rigid case.

Let $O$ be the center of the polhode for the elastic Earth (i.e., the average location of $R$ over a complete period of the wobble) and let us work in a geographic frame of reference, centered at $O$.  As shown in \Fig{fig:Newcomb}, $S$ describes its own circular trajectory, with fixed
\be
k_{\rm w} = OS/OR
\la{eq:kw}
\ee
related to the speed of rotation $\Omega$ and to the planet's elastic modulus.  The parameter $k_{\rm w}$ lies between between 0 (rigid case) and 1 (fluid case).

The centrifugal deformation preserves the Earth's ellipticity, leaving the magnitudes of $I$ and $I_3$ unaffected.  Newcomb's insight was that the planet's internal forces of cohesion will therefore make $R=(x,y)$ precess instantaneously about $S=(x_S, y_S)$ with the same $\omega_{\rm Eu}$ of \Eq{eq:Euler-f}:
\be
\left\{ \begin{array}{l}
\dot x = - \omega_{\rm Eu} \left( y - y_S \right) \\
\dot y = \omega_{\rm Eu} \left( x - x_S \right) \end{array} \right. ~.
\la{eq:Newcomb-Eu}
\ee
By \Eq{eq:kw}, $x_S = k_{\rm w} x$ and $y_S = k_{\rm w} y$, so that \Eq{eq:Newcomb-Eu} reduces to
\be
\ddot x + \omega_{\rm Ch}^2 \, x = \ddot y + \omega_{\rm Ch}^2 \, y = 0 ~,
\la{eq:Newcomb-x}
\ee
with
\be
\omega_{\rm Ch} = \omega_{\rm Eu} \left(1 - k_{\rm w} \right)
\la{eq:Newcomb-Ch}
\ee
the angular velocity of the precession of $R$ around $O$.  The parameter
\be
k_{\rm w} = 1 - \frac{\omega_{\rm Ch}}{\omega_{\rm Eu}} = 1 - \frac{306}{433} = 0.293
\la{eq:Love}
\ee
measures the Earth's deformability and is consistent with the Love number $k_2$ computed from the magnitudes of terrestrial and oceanic tides. \cite{Lambeck,Stacey}

\begin{figure}[b]
\sidecaption
\includegraphics[width=.45\textwidth]{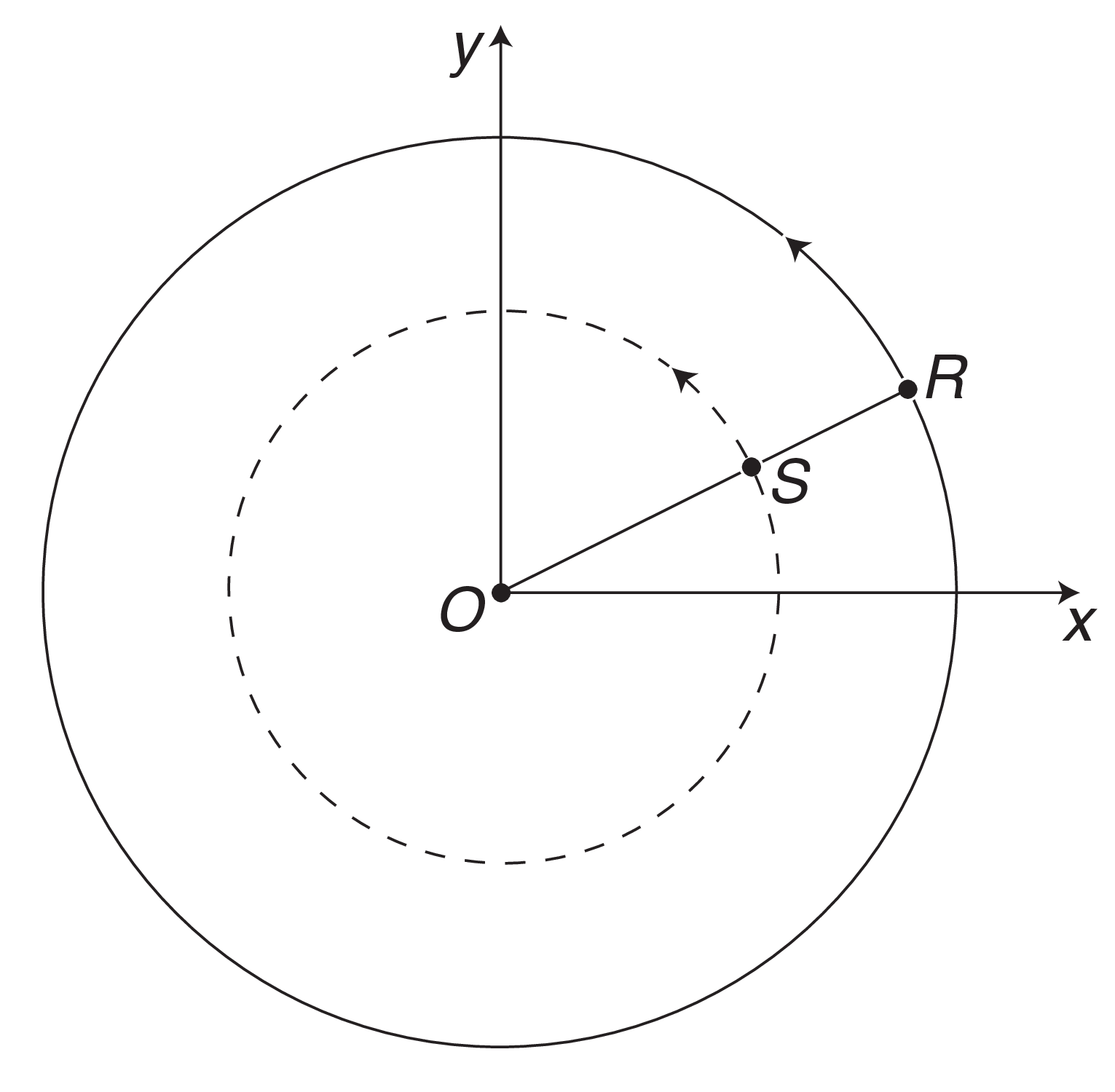}
\caption{Free precession of a deformable Earth.  $R$ is the instantaneous North Pole and $S$ is the symmetry pole.  The geographic coordinates $(x,y)$ are centered at $O$, which corresponds to the average position of $R$.}
\label{fig:Newcomb} 
\end{figure}

This simple analysis reveals that the dynamics of a deformable spinning body is subject to a feedback: the displacement of $R$ (the wobble) affects the displacement of $S$ (due to centrifugal deformation), while the displacement of $S$ (the deformation) in turn affects the displacement of $R$ (due to precession).  If some of the mass of the planet is in the form of circulating fluids (with their own kinetic energy in the geographic reference frame), we shall see that this feedback can be positive, destabilizing the wobble-less equilibrium.

%%%%%%%%%%%%%%%%%%%%%%%%%%%%%%%%%%%%%%%%%%%%%%%%%%%%%%%%%%%%%%%
\section{Dissipation and maintenance}
\label{sec:dissipation}

In the body frame (see \Fig{fig:Euler}), we write
\be
\vv \Omega = \left( \Omega_1, \Omega_2, \Omega_3 \right)
\la{eq:Omega}
\ee
with $\Omega_\perp \equiv \sqrt{\Omega_1^2 + \Omega_2^2}$.  The magnitude squared of the angular momentum $\vv M$ is therefore
\be
M^2 = I^2 \Omega_\perp^2 + I_3^2 \Omega_3^2 ~.
\la{eq:J}
\ee
Internal forces keep $M$ fixed, as they produce no net torque.  The kinetic energy of rotation is
\be
E = \frac{1}{2} \left( I \Omega_\perp^2 + I_3 \Omega_3^2 \right)  = E_0 + \frac{I}{2} \left( 1 - \frac{I}{I_3} \right) \Omega_\perp^2 ~,
\la{ec:E}
\ee
where $E_0\equiv M^2 / 2 I_3$ is the minimum energy consistent with conservation of $M$.

Let $\alpha = \arcsin (\Omega_\perp / \Omega)$ be the angle between $\vv \Omega$ and $\vh{x}_3$, in radians.  For small $\alpha$ we may express the wobble's energy as
\be
E_{\rm w} = \frac{I}{2} \left( 1 - \frac{I}{I_3} \right) \Omega_\perp^2  \simeq  \frac{I}{2} \left( 1 - \frac{I}{I_3} \right) \Omega^2 \alpha^2 ~.
\la{eq:Ew}
\ee
For the Earth this gives $E_{\rm w} \sim \alpha^2 \times 10^{27}$ J.

Displacement of the equatorial bulge (and therefore of the symmetry pole $S$ in \Fig{fig:Newcomb}) induces a small variation in ocean levels, with the same period as the wobble's.  Friction as this ``pole tide'' moves through straits and shallow oceans dissipates $E_{\rm w}$.  Another source of dissipation is the anelasticity of molten material under the Earth's crust.  We would therefore expect the amplitude of the free wobble to decay as
\be
\alpha (t) = e^{-t/\tau} \alpha_0
\la{eq:decay}
\ee
with $\alpha_0$ set by the Earth's state of rotation at the time $t=0$ when it solidified, and $\tau$ a time constant small compared to the age of the Earth.  In terms of the quality factor
\be
Q_{\rm w} \equiv \frac{\tau \omega_{\rm Ch}}{2} ~,
\la{eq:Qw}
\ee
the power that the wobble dissipates is
\be
P_{\rm w} = \frac{E_{\rm w} \, \omega_{\rm Ch}}{Q_{\rm w}} \sim \frac{\alpha^2}{Q_{\rm w}} \times 10^{20} ~ \hbox{W} ~.
\la{eq:Pw}
\ee
The fact that $\alpha$, averaged over many periods of the wobble, remains of order $10^{-6}$ indicates that some mechanism injects into it an average power of order $Q_{\rm w}^{-1} \times 10^{8}$ W, compensating for the dissipated $P_{\rm w}$. \cite{Lambeck,Stacey}

Stochastic perturbations to the Earth's mass distribution, such as might result from seismic events, could cause random displacements of $\vh{x}_3$, preventing it from becoming aligned with $\vv \Omega$.  Jeffreys developed a method for estimating $Q_{\rm w}$ from the data, assuming that the wobble is re-excited stochastically \cite{Jeffreys1}.  Jeffreys and others have thus obtained estimates for $Q_{\rm w}$ ranging from 37 to 1,000 \cite{Wilson}.  These estimates are not only uncertain, they generally require more dissipation than can be accounted for by known mechanisms of tidal friction and mantle anelasticities \cite{Jeffreys1,Lambeck,Stacey}, while the corresponding $P_{\rm w}$ seems too high to be compensated by seismic activity \cite{Souriau,Chao}.  Moreover, there is statistical evidence that the Chandler wobble is predominantly a deterministic process. \cite{Frede1,Frede2,Frede3}

Some authors have therefore concluded that the wobble is forced by geophysical fluid circulations \cite{Plag,Gross,Aoyama1,Aoyama2}.  In that case the estimates of $Q_{\rm w}$ obtained by Jeffreys's method do not apply directly (a point stressed in \cite{Lambeck}, but ignored by some authors).  The key difficulty with such models is that a damped and forced linear oscillator ends up moving at the forcing frequency.  Even in parametric or non-linear resonances, the frequency of steady oscillation is usually a rational multiple $a/b$ of the forcing frequency, with integer $a$ and $b$ of low order \cite{Landau}.  In fact, the polar motion has a significant component with a 12-month period, presumably forced by meteorological and other mass transfers connected with the seasons.  This seasonal effect is subtracted from the data to isolate the Chandler wobble \cite{Klein,Lambeck,Stacey}.  The polar motion's power spectrum is inconsistent with a high-quality oscillation ($Q \gtrsim 300$) being driven by an external power spectrum that is smooth in the vicinity of $\omega_{\rm Ch}$ \cite{Mandelbrot}.  The observed wobble must therefore be forced by an external power spectrum peaked around the resonant $\omega_{\rm Ch}$, which seems like an implausible coincidence. \cite{Lambeck,Stacey}

Moreover, it is well known that in the 1920s the amplitude of the Chandler wobble decreased sharply, only to re-start with a phase jump of nearly $180^\circ$ \cite{Lambeck}.  A recent analysis finds two other brief extinctions, followed by phase jumps, in the 1850s and 2000s \cite{Malkin}.  These irregularities are difficult to explain in stochastic re-excitation or forced resonance models, as they are not associated with obvious geophysical events.

%%%%%%%%%%%%%%%%%%%%%%%%%%%%%%%%%%%%%%%%%%%%%%%%%%%%%%%%%%%%%%%
\section{Self-oscillation and intermittence}
\label{sec:Hopf}

Consider the portion of the Earth's mass made of circulating fluids.  In the body frame of the solid Earth, these circulations carry significant kinetic energy but negligible net angular momentum.  Resistance to displacing geographically the mean route of a circulation may be interpreted as an effective rigidity.  With respect to the Earth's rotation, we therefore express the total tensor of inertia as 
\be
I^{\rm tot}_{ij} = I_{ij} + \varepsilon \tilde{I}_{ij} ~,
\la{eq:tensor}
\ee
with $I_{ij}$ given by the distribution of the mass at rest with respect Earth's solid surface, and $\tilde{I}_{ij}$ by flows whose distribution is expressed in Eulerian coordinates (fixed geographically), rather than in Lagrangian coordinates associated to individual mass elements (see \cite{Tritton}).

The moments of $I_{ij}$ and $\tilde{I}_{ij}$ in \Eq{eq:tensor} are taken to be of the same order, with $\varepsilon$ indicating the small fraction that the circulations contribute to the moments of $I^{\rm tot}_{ij}$.  Euler's equations for the free top may be expressed as
\be
\frac{d' \vv M}{dt} + \vv \Omega \times \vv M = 0 ~,
\la{eq:Euler-M}
\ee
where $d'/dt$ measures velocity in the body frame \cite{Landau}.  For $M_i =  \sum_j (I_{ij} + \varepsilon \tilde{I}_{ij}) \Omega_j$, \Eq{eq:Euler-M} implies that the equation of motion for the free precession of $\vv \Omega$ is the weighted superposition of the equations of motion for the precessions separately induced by $I_{ij}$ and $\varepsilon \tilde{I}_{ij}$.

Let $S$ be the symmetry pole for $I_{ij}$.  Let us assume, for simplicity, that $\tilde{I}_{ij}$ has its own symmetry pole $\tilde{S}$, as shown in \Fig{fig:powered}.  For an asymmetric $\tilde{I}_{ij}$, we would consider precession about the principal axis of greatest moment of inertia.  This precession would not be uniform \cite{Landau}, but our conclusions would be essentially unaffected.

The centrifugal deformation of $I_{ij}$ propagates with the speed of sound in the relevant material, which is very fast compared to the wobble, so that the adjustment of $S$ to the displacement of $R$ is nearly instantaneous \cite{Klein}.  On the other hand, the adjustment of $\tilde{S}$ can show an appreciable delay, caused by the circulating fluids' inertia.  For $R=(x, y)$, we therefore take 
\be
\tilde{S}(t) = \beta \left( x(t-c), y(t-c) \right) ~,
\la{eq:delay}
\ee
where $\beta \equiv O\tilde{S} / OR$ (\Fig{fig:powered} is drawn for constant $\beta$).  The precession of $R$ around $\tilde{S}$ would then be described by
\be
\left\{ \begin{array}{l}
\dot x(t) = - \tilde{\omega}_{\rm Eu} \left[ y(t) - \beta y(t-c) \right] \\
\dot y(t) = \tilde{\omega}_{\rm Eu} \left[ x(t) - \beta x(t-c) \right]
\end{array} \right. ~,
\la{eq:powered}
\ee
with $\tilde{\omega}_{\rm Eu} \equiv \tilde{\Omega}_3 (\tilde{I}_3 - \tilde{I}) / \tilde{I}$ given by the moments of inertia $\tilde{I}$ and $\tilde{I}_3$ for the tensor $\tilde{I}_{ij}$, and by the component $\tilde{\Omega}_3$ of the angular velocity along the axis of $\tilde{S}$.  In terms of $x$, \Eq{eq:powered} corresponds to
\be
\ddot x (t) = - \tilde{\omega}_{\rm Eu}^2 \left[ x(t) + 2 \beta x (t - c) - \beta^2 x (t - 2c) \right] ~,
\la{eq:powered-x}
\ee
and equivalently for $y$.

\begin{figure} [b]
\sidecaption
\includegraphics[width=.55\textwidth]{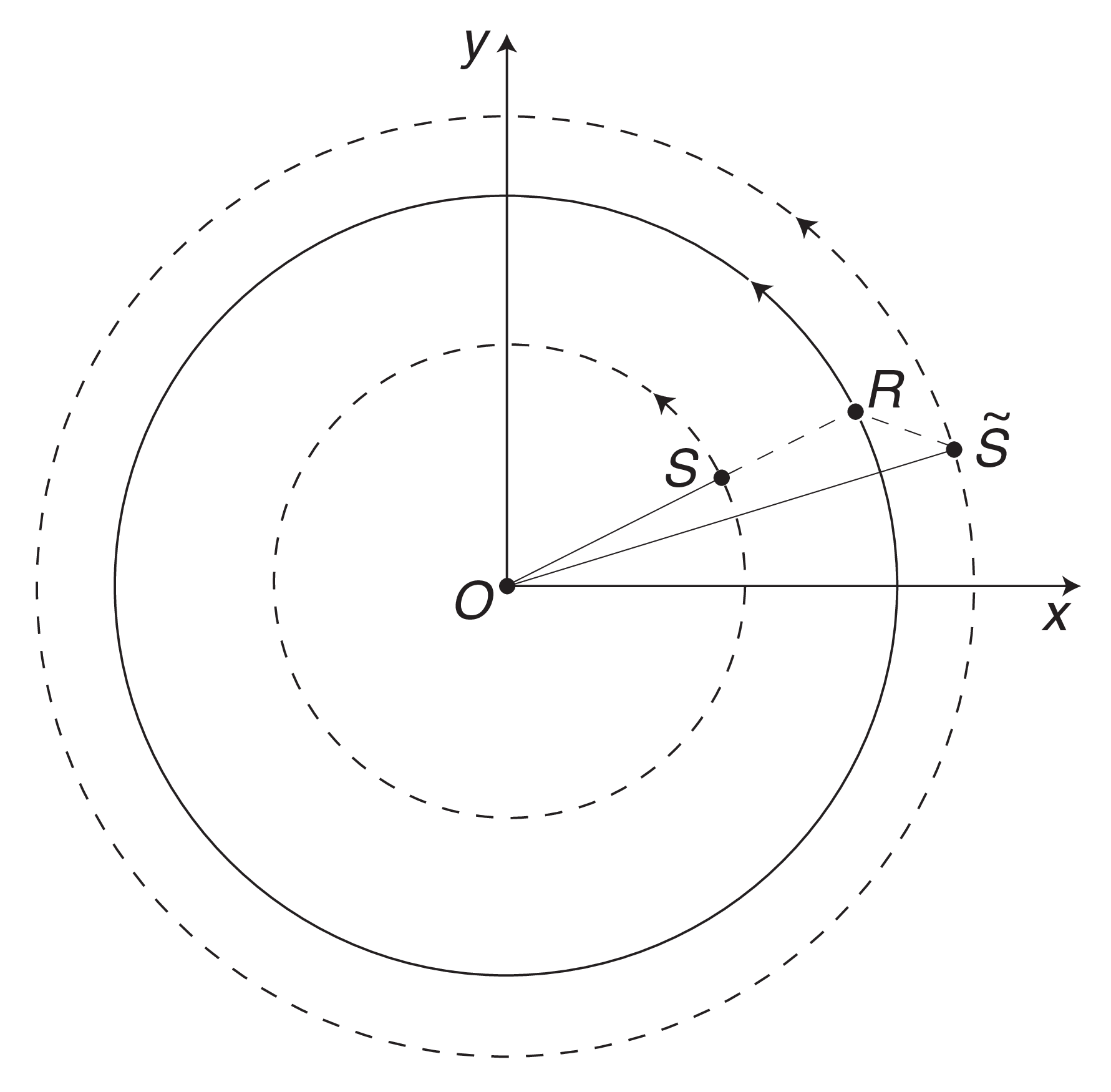}
\caption{\small Torqueless precession of a deformable Earth containing fluid circulations.  $R$ is the instantaneous North Pole, $S$ is the symmetry pole of the portion of the mass at rest with respect to the Earth's solid surface, and $\tilde{S}$ is the symmetry pole of the part of the mass in geophysical circulations. \la{fig:powered}}
\end{figure}

Superposing the precession about $\tilde{S}$, given by \Eq{eq:powered-x} and weighed by $\varepsilon \ll 1$, on the precession about $S$ given by \Eq{eq:Newcomb-x}, and taking \hbox{$c \ll \pi / \omega_{\rm Ch}$} (which lets us approximate $ x(t-c) \simeq x(t) -c \dot x (t)$ in the full equation of motion) we obtain
\be
\ddot x + 2 c \varepsilon \beta (1 - \beta) \tilde{\omega}_{\rm Eu}^2 \dot x + \omega_{\rm Ch}^2 x = 0 ~.
\la{eq:antidamped}
\ee
The coefficient of $\dot x$ in \Eq{eq:antidamped} is negative for $\beta > 1$, indicating that the wobble {\it absorbs} energy.  As long as
\be
\gamma \equiv 2 c \varepsilon \beta (\beta - 1) \tilde{\omega}^2_{\rm Eu} > \omega_{\rm Ch}/Q_{\rm w}
\la{eq:threshold}
\ee
this effect will dominate over viscous damping, causing $R$ to spiral away from $O$ until non-linear dissipative effects match the power input, stabilizing the wobble's amplitude. In the theory of ordinary differential equations, this corresponds to approaching a weakly non-linear limit cycle; see \cite{Jenkins}.  (Note that if the mass associated with $\tilde{S}$ did not have significant kinetic energy in the geographic frame, $c$ would be negligible and it would be impossible to maintain $\beta > 1$.)

The parameters $\varepsilon, \beta$ vary with the magnitude and distribution of the geophysical circulations.  The anelasticity of $\tilde{I}_{ij}$ implies that centrifugal deformation would tend to make $\beta \to 1$, killing the anti-damping in \Eq{eq:antidamped}.  However, natural changes in circulation patterns may counteract this, maintaining $\beta > 1$ over long periods (recall that $OR \sim 6$ m).

Instead of tending to a limit cycle, the wobble may start to decay if $\tilde{S}$ approaches close enough to the circle $OR$ for the condition of \Eq{eq:threshold} to fail.  Sensitivity to $\beta$, determined by the precise geographical distribution of the circulations, implies that extinctions of the wobble need not be associated with major geophysical irregularities.  As $OR \to 0$, $\beta \gg 1$ becomes more likely, and we therefore expect the wobble to turn on again, with a random phase jump related to its brief quiescent period.  The turning on or off of a self-oscillation as the net linear damping changes sign is known mathematically as ``Hopf bifurcation'' (see \cite{Jenkins}).

The conditions that the delay $c$ in \Eq{eq:delay} remain fixed and very small compared to the period of the wobble may be relaxed without qualitatively altering our conclusions.  Physically, the point is that the wobble itself induces a modulation of the geophysical circulations, corresponding to the precession of $\tilde{S}$ in \Fig{fig:powered}.  For $\beta > 1$ and \hbox{$0 < \omega_{\rm Ch} c < \arccos(1/\beta)$}, the resulting force on the solid Earth, described by \Eq{eq:powered-x}, {\it leads} the wobble, thus supplying to it net power (see \cite{Georgi}).

%%%%%%%%%%%%%%%%%%%%%%%%%%%%%%%%%%%%%%%%%%%%%%%%%%%%%%%%%%%%%%%
\section{Outlook}
\label{sec:outlook}
 
In the self-oscillatory model offered here, the Chandler wobble is maintained by a positive feedback between the wobble and the centrifugal deformation of the part of the Earth's mass in fluid circulations.  The energy to maintain the wobble comes from the flows, which take it in turn from the heat of solar radiation and of the Earth's internal radioactivity.  This is a picture of the wobble, not as an unphysical perpetual motion, but as a low-efficiency heat engine, with the geophysical fluids playing the part of the working substance and the solid Earth acting as the piston.  That heat engines and motors in general may be conceptualized as self-oscillators has been stressed in \cite{Andronov,Jenkins}.

Evidence of the modulation of the circulations may already have been detected in the atmosphere \cite{Plag,Aoyama1,Aoyama2}, requiring only re-interpreting that signal as resulting from the wobble's feedback, rather than being an intrinsic property of the circulations.  The positive local Lyapunov exponent that \cite{Frede1,Frede2,Frede3} find in the Chandler wobble data may reflect, not chaos (an aperiodic and strongly non-linear phenomenon), but rather the linear anti-damping of \Eq{eq:antidamped} (which makes the wobble-less equilibrium unstable).  The reported variability of this local Lyapunov exponent, and its tendency to increase when the El Ni\~no Southern Oscillation (ENSO) is on its maximum (El Ni\~no) or minimum (La Ni\~na) temperature phase, seem consistent with our model, as ENSO extrema may be associated with stronger atmospheric currents around the equatorial region, and therefore with an increased $\varepsilon$ in \Eq{eq:tensor}.

This simple model raises issues that may be relevant beyond geophysics.  A deterministic self-oscillation turned on and off by a stochastic parameter may lead to behaviors difficult to capture with simple statistical tools.  (For a review of how heavy-tailed distributions may emerge when a stochastic parameter triggers repeated dynamical bifurcations, see \cite{Volchenkov}.)  This model of the Chandler wobble may therefore offer a case study on the importance of better understanding the relations between stochastic and deterministic dynamics in complex systems. \\

%%%%%%%%%%%%%%%%%%%%%%%%%%%%%%%%%%%%%%%%%%%%%%%%%%%%%%%%%%%%%%

\begin{acknowledgement}
The author thanks Eric Alfaro, Jorge Amador, and Paul O'Gorman for discussions on meteorological issues, as well as Howard Georgi and Jos\'e Gracia-Bond\'ia for encouragement and advice on this project.
\end{acknowledgement}

\input{referenc}

\end{document}

%% file: referenc.tex
%%%%%%%%%%%%%%%%%%%%%%%% referenc.tex %%%%%%%%%%%%%%%%%%%%%%%%%%%%%%
% sample references
% %
% Use this file as a template for your own input.
%
%%%%%%%%%%%%%%%%%%%%%%%% Springer-Verlag %%%%%%%%%%%%%%%%%%%%%%%%%%
%
% BibTeX users please use
% \bibliographystyle{}
% \bibliography{}
%